\documentclass[prb,twocolumn,amsmath,amssymb,floatfix,eqsecnum]{revtex4}
\usepackage{graphicx}
\usepackage{bm}

\begin{document}


\title{
Strongly coupled quantum criticality with a Fermi surface in two
dimensions: fractionalization of spin and charge collective modes}

\author{Subir Sachdev}
\email{subir.sachdev@yale.edu}
\homepage{http://pantheon.yale.edu/~subir}
\affiliation{Department
of Physics, Yale University, P.O. Box 208120, New Haven CT
06520-8120}
\author{Takao Morinari}
\email{takao.morinari@yale.edu}
\altaffiliation[Permanent address:
]{Yukawa Institute for Theoretical Physics, Kyoto University Kyoto
606-8502, Japan} \affiliation{Department of Physics, Yale
University, P.O. Box 208120, New Haven CT 06520-8120}

\date{July 4, 2002}

\begin{abstract}
We describe two dimensional models with a metallic Fermi surface
which display quantum phase transitions controlled by strongly
interacting critical field theories below their upper critical
dimension. The primary examples involve transitions with a
topological order parameter associated with dislocations in
collinear spin density wave (``stripe'') correlations: the
suppression of dislocations leads to a fractionalization of spin
and charge collective modes, and this transition has been proposed
as a candidate for the cuprates near optimal doping. The coupling
between the order parameter and long-wavelength volume and shape
deformations of the Fermi surface is analyzed by the
renormalization group, and a runaway flow to a non-perturbative
regime is found in most cases. A phenomenological scaling analysis
of simple observable properties of possible second order quantum
critical points is presented, with results quite similar to those
near quantum spin glass transitions and to phenomenological forms
proposed by Schr\"oder {\em et al.} (Nature {\bf 407}, 351
(2000)).
\end{abstract}

\maketitle

\section{Introduction}
\label{sec:intro}

An important property of most quantum phase transitions studied to
date in systems with a metallic Fermi surface in spatial
dimensions $d \geq 2$ is that the critical field theory for the
order parameter is a free Gaussian theory \cite{maki,hertz}. This
result has its origin in the fact that the order parameters
considered can be expressed as a fermion bilinears, and
consequently the order parameter fluctuations are efficiently
overdamped and suppressed by fermionic particle and hole
excitations near the Fermi surface. The temperature dependencies
of physical observables near the quantum critical point have been
perturbatively computed
\cite{mathon,tvr,moriya,lonzarich,millis1,millis2,scs,vadim,belitz},
and these can be understood as corrections to
scaling\cite{millis1} at the Gaussian critical point\cite{foot}.

The quantum and thermal fluctuations near quantum phase
transitions in Fermi systems have been used to interpret
experiments in a variety of correlated electron materials. In the
cuprate superconductors, the anomalous properties in the normal
state have been described in terms of the proximity to a quantum
phase transition associated with the onset of spin density wave
(SDW) or charge density wave (CDW) order in a Fermi liquid
\cite{andrey,monthoux,rome}. However, the observed anomalous
properties extend to quite high temperatures, and it would be
preferable to explain these in theories with stronger
non-linearities among the order parameter modes. Further, many of
the anomalous properties extend to the optimal doping regime where
there is no strong evidence of a large correlation length of the
SDW and CDW orders.

This paper describes an alternative class of quantum critical
points whose critical theories are non-Gaussian, and which remain
strongly coupled even in the presence of a Fermi surface because
they are below their upper critical dimension. Only in such
theories can the order parameter relaxation rate, and possibly the
quasiparticle energy width, be generically equal to $k_B T$ times
a universal numerical constant\cite{book} ($T$ is the absolute
temperature). Examples of such quantum phase transitions abound in
insulators and superconductors\cite{science,bfn,vzs}, and some of
these have been used to explain low temperature properties of the
cuprate superconductors \cite{sy,pg,dsz,zds}. This paper will
discuss transition in metals with a Fermi surface, and so the
possible regime of applicability is restricted to higher
temperatures where superconductivity is absent.

Our primary focus will be on a transition with a non-local
``topological'' order associated with certain defects in the
SDW/CDW order. However, long-range SDW or CDW order will not be
present on either side of the critical point. This transition
(shown in Fig.~\ref{phase} below) is being offered as a
candidate\cite{zaanen,zds} for a possible optimal doping quantum
critical point \cite{tallon,valla} in the cuprate superconductors.
This proposal for the optimal doping quantum critical point must
be distinguished from a conventional SDW ordering transition in a
{\em superconductor\/} discussed recently by one of
us\cite{dsz,zds}; the latter transition occurs at lower doping
concentrations (near 1/8) and was used to predict and explain a
number of recent {\em low temperature\/} neutron scattering and
STM experiments. Here we are interested in higher dopings and
temperatures, where, as we have already noted, the correlation
length of conventional SDW/CDW order is surely quite small, but,
as we shall discuss below, correlations in a topological order may
be far more robust.

We also hope for applications of our theory in other correlated
materials, like the heavy fermion systems, where many experiments
are in disagreement with the conventional weakly coupled
theories\cite{piers,expts}. In particular, we will show that the
strongly-coupled critical points predict a scaling structure for
physical observables which is quite similar to those proposed by
Schr\"oder {\em et al.} \cite{almut} Related scaling structures
are also known to appear near strongly-coupled spin glass quantum
critical points, and this is reviewed in Appendix~\ref{app:qsg}.

There is one simple route to obtaining a strongly coupled field
theory in the presence of a Fermi surface that has been discussed
earlier \cite{scs}: use restrictions from momentum conservation to
prevent the order parameter from coupling linearly to low energy
fermionic excitations. For example, a transition with the onset of
SDW order at the wavevector ${\bf K}$, but the geometry of the
Fermi surface such that no two points on the Fermi surface are
separated by ${\bf K}$. In such a situation the order parameter
excitations remain undamped, and the fermions appear to be
innocuous bystanders to the transition. However, it turns out that
the Fermi surface can still be relevant for the structure of the
critical theory in certain situations; these effects are related
to those that arise in the transitions we do discuss below, and so
we defer discussion of this point until Section~\ref{sec:sdw}.

Turning to transitions characterized by a non-local, `topological'
order parameter, a situation that has been much discussed in the
recent literature\cite{Senthil99} has the electron,
$c^{\dagger}_\varsigma$, ($\varsigma = \uparrow, \downarrow$ is a
spin index) fractionalize into a spinless charge $e$ boson, $b$,
and a neutral spin 1/2 fermion $f_\varsigma$: $
c^{\dagger}_\varsigma = b f_\varsigma^{\dagger} $ This
fractionalization transition is described by a $Z_2$ gauge theory
\cite{z2}, and both $b$ and $f_\varsigma$ carry a unit gauge
charge; the physical fermion, $c^{\dagger}_\varsigma$, is, of
course, gauge neutral. If we denote the $Z_2$ gauge field by
$\sigma_{ij}$ ($i,j$ are site labels), then terms like $
\sigma_{ij} f_{i \uparrow} f_{j \downarrow} $ will generically
appear in the effective Hamiltonian \cite{Senthil99}. In the
fractionalized phase, where the fluctuations of $\sigma_{ij}$ are
effectively quenched, this term is a BCS-like pairing of the
$f_\varsigma$ fermions, and it implies that the Fermi surface will
generically be gapped out (except possibly at special points if
the pairing is anisotropic). Consequently, this fractionalization
transition does not provide us with a candidate quantum phase
transition in the presence of a Fermi surface, and we will not
consider it further in this paper.

The main focus of this paper shall be on a fractionalization
transition that does less damage to the integrity of the electron:
the electron (and hence its Fermi surface) remains intact on both
sides of the transition, but charge neutral collective modes (in
the particle-hole sector) do fractionalize; only at the quantum
critical point, and in the associated non-zero temperature quantum
critical region, is the electronic quasiparticle ill-defined.
Zaanen {\em et al.} \cite{zaanen} have recently given an appealing
pictorial description of such a transition. A transition with
closely related physical content was introduced in
Ref.~\onlinecite{zds}; we will follow the latter approach here,
and will describe the physical content and experimental motivation
below in Section~\ref{sec:op}. There have also been earlier
discussions of the fractionalization of order parameters in other
physical contexts \cite{Lammert93,bookfrac,Zhou}.

After introducing the order parameters in Section~\ref{sec:op}, we
write down and classify quantum field theories for the critical
points in Section~\ref{sec:qft}. We pause briefly in
Section~\ref{sec:sdw} to consider the conventional transitions
mentioned above, in which the fermion bilinear order parameter is
not damped by Fermi surface excitations because of restrictions
arising from momentum conservation, and show that it is described
by field theories similar to those in Section~\ref{sec:qft}. The
latter are subjected to a renormalization group analysis in
Section~\ref{sec:rg}, and the physical implications of the results
are noted in Section~\ref{sec:phys}.

\section{Order parameter and physical motivation} \label{sec:op}

Consider a correlated electronic system in two dimensions with
enhanced SDW correlations at the wavevectors ${\bf K}_x$ and ${\bf
K}_y$; these are vectors of equal length pointing along the $x$
and $y$ axes respectively. Similar considerations apply to other
choices for the ordering wavevectors, but we will focus on this
case because it directly applies to the doped cuprates
\cite{wakimoto} at doping concentrations above $0.055$. We can
write for the spin operator $S_{\alpha} ({\bf r}, \tau)$ (${\bf
r}$ is a spatial co-ordinate, $\alpha = x,y,z$ labels spin
components, and $\tau$ is imaginary time):
\begin{equation}
S_{\alpha} ({\bf r}, \tau) = \mbox{Re} \left[e^{i {\bf K}_{x}
\cdot {\bf r}} \Phi_{x \alpha} ({\bf r}, \tau)+  e^{i {\bf K}_{y}
\cdot {\bf r}} \Phi_{y \alpha}({\bf r}, \tau)\right]; \label{e1}
\end{equation}
$\Phi_{x\alpha}$ and $\Phi_{y\alpha}$ are then the SDW order
parameters. Except of the case of two sublattice order (${\bf K}_x
= (\pi,0)$ or $(\pi,\pi)$), these order parameters are complex
numbers. Concommitant with this SDW correlations, symmetry demands
\cite{zachar,lt23} that there are also ``bond order'' correlations
at half the wavelength:
\begin{eqnarray}
Q_{{\bf a}} ({\bf r}, \tau) &\equiv& S_{\alpha} ({\bf r}, \tau)
S_{\alpha} ({\bf r} + {\bf a}, \tau) \nonumber \\
&\approx & \mbox{Re} \left[ e^{2 i {\bf K}_{x} \cdot {\bf r} + i
{\bf K}_x \cdot {\bf a}} \Phi_{x \alpha}^2 ({\bf r}, \tau) \right]
+ \ldots \label{e2}
\end{eqnarray}
there is an implied summation over the repeated spin index
$\alpha$. The vector ${\bf a}$ represents a bond (say the nearest
neighbor vector), we have assumed that $\Phi_{x\alpha}$ does not
vary significantly over the spatial distance ${\bf a}$, and the
ellipses denote numerous other similar terms which can be deduced
from (\ref{e1}). For ${\bf a}$ a nearest-neighbor vector, $Q_{{\bf
a}}$ measures the modulations in the exchange energy, while for
${\bf a} = 0$, $Q_{{\bf a}}$ measures the local charge density
(for the $t$-$J$ model the charge density is linearly related to
the on-site $S_{\alpha}^2$). So, quite generally, the bond order
parameters, $\Phi_{x,y \alpha}^2$, measure modulations in all
observables invariant under spin rotations and time reversal (such
the electron kinetic, pairing, or exchange energies). The
modulation in the total charge density may be strongly suppressed
by the long-range Coulomb interactions, but this suppression is
not expected to apply to observables with ${\bf a } \neq 0$. We
will loosely refer to such bond order modulations as CDW order in
this paper, but the caveats discussed in this paragraph must be
kept in mind.

Now examine the structure of the effective potential, $V
(\Phi_{x\alpha})$ controlling fluctuations of $\Phi_{x \alpha}$
(parallel considerations apply implicitly in the remainder of this
subsection to $\Phi_{y \alpha}$). On general symmetry grounds
\cite{zachar,zds}, this potential has the form
\begin{eqnarray}
V (\Phi_{x \alpha}) = \widetilde{s}  \Phi_{x \alpha}^{\ast}
\Phi_{x \alpha}^{\ast} &+& \frac{\widetilde{u}}{2} \Phi_{x
\alpha}^{\ast}
\Phi_{x \beta}^{\ast} \Phi_{x \alpha} \Phi_{x \beta} \nonumber \\
&+& \frac{\widetilde{v}}{2} \Phi_{x \alpha}^{\ast} \Phi_{x
\alpha}^{\ast} \Phi_{x \beta} \Phi_{x \beta} + \ldots \label{vphi}
\end{eqnarray}
where $\widetilde{s}$, $\widetilde{u}$, $\widetilde{v}$ are
phenomenological Landau parameters, and there is again an implied
summation over repeated spin indices $\alpha,\beta$. In the usual
Landau theory, the transition to the onset long-range SDW order
happens when $\widetilde{s}$ first becomes negative, and the
minimum of the potential moves from $\Phi_{x \alpha} = 0$ to a
non-zero value. The theory of the critical point then accounts for
fluctuations of $\Phi_{x \alpha}$ about the value of $\Phi_{x
\alpha}=0$, and these allow for strong fluctuations in both the
amplitudes and the phases of the three complex fields $\Phi_{x
\alpha}$.

In much of the discussion on stripe physics in the literature (see
{\em e.g.} Ref.~\onlinecite{zaanen}), the physical picture
advocated for the onset of long range SDW/CDW (or ``stripe'')
order is quite different. As one decreases the doping from a Fermi
liquid towards a Mott insulator, initial local stripe-like
correlations form, the the orientation and phase of these regions
changes from point to point; only when these locally ordered
regions align with each other, is there long range SDW/CDW order.
In terms of our order parameter formulation, this means that
locally the amplitude $\Phi_{x \alpha}^{\ast} \Phi_{x \alpha} $
first becomes large, and that long range order develops
subsequently upon alignment of the orientation and phases of
$\Phi_{x \alpha}$.

In the familiar $\varphi^4$ field theory of a {\em real} vector
order parameter, this change in perspective from a ``soft-spin''
Landau theory (with strong amplitude fluctuations) to a
``hard-spin'' perspective (with amplitude fluctuations quenched)
does not make a fundamental difference in the physics being
considered. The orientation fluctuations in the hard-spin approach
are described by a non-linear $\sigma$ model, whose critical
properties can be analyzed by expansion away the lower critical
dimension. Conversely, the soft-spin approach leads to an
expansion away from the upper critical dimension. However, these
are merely complementary approaches to the same critical point,
and the quantum numbers of the excitations in the phases flanking
the critical point are the same in both approaches; indeed a $1/N$
expansion ($N$ is the number of field components) can interpolate
between the soft-spin and hard-spin theories, and also between the
upper and lower critical dimensions \cite{bz}.

For the present situation with a {\em complex} vector order
parameter the situation is dramatically different, and there is a
fundamental difference between the soft-spin and hard-spin
approaches to fluctuations implied by the effective potential
(\ref{vphi}). The key difference is that not all order parameter
configurations with a fixed $\Phi_{x \alpha}^{\ast} \Phi_{x
\alpha}$ are physically equivalent, and they cannot all be rotated
into each other by a symmetry transformation. A minimization of
(\ref{vphi}) at fixed $\Phi_{x \alpha}^{\ast} \Phi_{x \alpha}$
shows two distinct classes of minima, chosen by the sign of
$\widetilde{v}$; their most general form is:
\begin{eqnarray}
({\rm I})~\widetilde{v}>0~&&:~~\Phi_{x \alpha} = n_{1\alpha} + i
n_{2 \alpha} \nonumber \\ && \mbox{with $n_{1,2\alpha}$ real,
$n_{1 \alpha}^2 = n_{2 \alpha}^2$ and $n_{1
\alpha} n_{2 \alpha} = 0$} \nonumber \\
({\rm II})~\widetilde{v}<0~&&:~~\Phi_{x \alpha} = e^{i \theta_x}
n_{x\alpha} \nonumber
\\ && \mbox{with $n_{x \alpha}$ real}
\label{ab}
\end{eqnarray}
By inserting into (\ref{e1}), it is easy to see that (I) describes
a spiral SDW for which the charge order in (\ref{e2}) vanishes,
while (II) is a collinear SDW with a modulation in the length of
the spin as a function of ${\bf r}$, and a concommitant charge
order. All experimental indications \cite{tran,ylee} are that the
spin fluctuations in the cuprates are always of the form (II). It
therefore seems physically reasonable to impose the amplitude
constraint implied by (II) at an early stage, and to examine the
theory associated with $V(\Phi_{x \alpha})$ in the limit of large
and negative $\widetilde{v}$ (stability requires
$\widetilde{u}+\widetilde{v}>0$). This last constraint is the more
formal statement of the physical requirement that the spin and
charge ordering is ``stripe-like''.

\begin{figure}
\centerline{\includegraphics[width=2.5in]{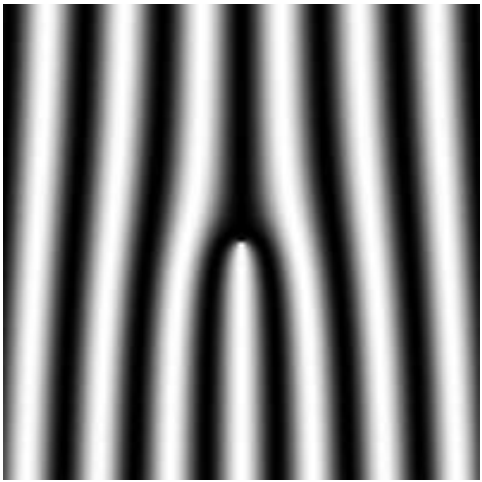}}
\caption{A stripe dislocation: gray scale plot of $\delta \rho
({\bf r}) \sim \cos(2 {\bf K}_x \cdot {\bf r} + 2\theta_x)$ (see
(\protect\ref{e2}) and (\protect\ref{ab})) with $\theta_x = \pi/2+
(1/2)\tan^{-1} (y/x)$ containing a half vortex. As noted below
(\protect\ref{e2}) $\delta \rho ({\bf r})$ should not be literally
interpreted as the electron charge density, and the modulation
about could be associated with {\em e.g.} the exchange per link.
The quantum phase transitions discussed in this paper involve a
transition from a Fermi liquid state in which these dislocations
proliferate to one in which they are suppressed; there is no
long-range spin or charge density wave order in either state (see
also Fig.~\protect\ref{phase}).} \label{dislocation}
\end{figure}

As has been discussed in Ref.~\onlinecite{zds}, the imposition of
the limit $\widetilde{v} < 0$ at the outset leads to a theory for
phase and orientation fluctuations with a non-trivial structure
that is very naturally described by a $Z_2$ gauge theory. With the
parameterization (II) in (\ref{ab}), and for fixed $\Phi_{x
\alpha}^{\ast} \Phi_{x \alpha}$, the order parameter $\Phi_{x
\alpha}$ takes values in the space $(S_2 \times S_1)/Z_2$, where
the $S_2$ pertains to the orientation of $n_{x \alpha}$, the $S_1$
is the phase factor \cite{commensurate} $e^{i \theta_x}$, while
the all-important $Z_2$ quotient accounts for the invariance of
$\Phi_{x \alpha}$ under $n_{x \alpha} \rightarrow -n_{x \alpha}$
and $\theta_x \rightarrow \theta_x + \pi$. We can define global
actions \cite{zds} for single-valued fields $n_{x \alpha}$ and
$e^{i \theta_x}$ only at the cost of introducing a $Z_2$ gauge
field $\sigma_{ij}= \pm 1$. A simple phenomenological form for the
spatial terms in such an action is \cite{zds}:
\begin{eqnarray}
\mathcal{S}_{Z_2} &=& -\frac{1}{g_{\theta}} \sum_{\langle ij
\rangle} \sigma_{ij} \cos(\theta_{xi} - \theta_{xj}) -
\frac{1}{g_n} \sum_{\langle ij \rangle} \sigma_{ij} n_{x\alpha i }
n_{x\alpha j} \nonumber \\
&+& K \sum_{\square} \prod_{\square} \sigma_{ij}, \label{z2flux}
\end{eqnarray}
where $i,j$ are lattice sites, and $g_{\theta}$, $g_n$, and $K$
are coupling constants that vary with doping. The $Z_2$ gauge
invariance ensures that all physical properties are invariant
under the $Z_2$ gauge transformation $\theta_{xi} \rightarrow
\theta_{xi} + (1-\sigma_i)\pi/2$, $n_{x\alpha i} \rightarrow n_{x
\alpha i} \sigma_i$ and $\sigma_{ij} \rightarrow \sigma_i
\sigma_{ij} \sigma_j$. The point defects associated with order
parameter space $(S_2 \times S_1)/Z_2$ were analyzed in
Ref.~\onlinecite{zds}, and a fundamental role is played by the
``stripe dislocation'', sketched in Fig~\ref{dislocation}, which
is a half-vortex in the angular field $\theta_x$.
This vortex also carries a $Z_2$ gauge flux of $-1$, as is easily
seen from (\ref{z2flux}).

We are now in a position to introduce our candidate transition for
strongly coupled quantum criticality in the presence of a Fermi
surface. Begin in a Fermi liquid state, with only short range spin
and charge order correlations: $\langle \Phi_{x \alpha} \rangle =
0$, $\langle \Phi_{x \alpha}^2 \rangle = 0$ ($\widetilde{s}$ in
(\ref{vphi}) is large and positive, $g_\theta$, $g_n$ in
(\ref{z2flux}) are not large) and only a small core energy for the
stripe dislocations, which therefore proliferate. The dislocation
core energy is controlled by the Maxwell term for the $Z_2$ gauge
field, $K$, and a small value of $K$ implies that the $Z_2$ gauge
field is in its confining state. In this state, the strong
fluctuations of the $Z_2$ gauge field ``bind'' the $n_{x\alpha}$
and $e^{i\theta_x}$ fluctuations together, and the appropriate
collective excitation is the conventional field $\Phi_{x \alpha}$
itself, as it appears in (\ref{e1}). The single $\Phi_{x \alpha}$
collective mode therefore describes both the SDW fluctautions (via
\ref{e1})) and the bond order/CDW fluctuations (via (\ref{e2})).

Now increase the dislocation core energy by increasing $K$, while
keeping $\widetilde{s}$ reasonably large (or $g_\theta$, $g_n$
small): this will decrease the number of stripe dislocations but
still maintain $\langle \Phi_{x \alpha} \rangle = 0$, $\langle
\Phi_{x \alpha}^2 \rangle = 0$. In the cuprates we imagine that
the increase in $K$ is associated with the decrease in carrier
concentration towards the Mott insulator. At a critical value of
$K_c$, the $Z_2$ gauge theory undergoes a transition into a
deconfined state in which the stripe dislocations are suppressed.
The transition at $K=K_c$ is the focus of interest in this paper.
In this deconfined state, the $n_{x \alpha}$ and $e^{i \theta_x}$
fluctuations become independent collective excitations: {\em
i.e.\/} the single spin/charge collective mode, $\Phi_{x \alpha}$,
has {\em fractionalized} into two independent modes. The bond
order, (\ref{e2}), fluctuations are now given by the `square' of
the $e^{i \theta}_x$ collective mode, while the spin fluctuations,
(\ref{e1}), are the product of the $n_{x \alpha}$ and $e^{i
\theta_x}$ collective modes.

\begin{figure}
\centerline{\includegraphics[width=3in]{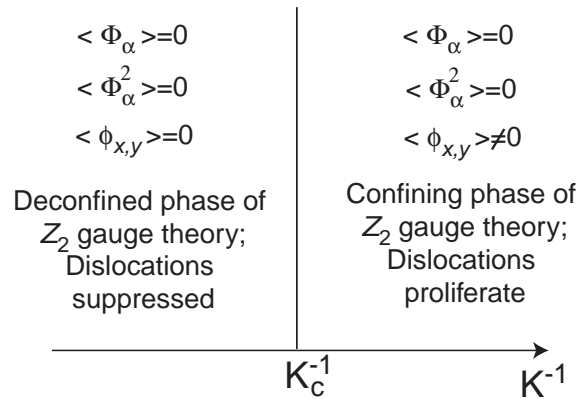}} \caption{The
quantum phase transition at $K=K_c$ is the one studied in this
paper. The phases on both sides of $K_c$ have well defined
electronic quasiparticles near a Fermi surface and no long range
spin or charge order ($\langle \Phi_{x,y \alpha} \rangle = 0$,
$\langle \Phi_{x,y \alpha}^2 \rangle = 0$). The order parameters
$\phi_{x,y}$ are defined in Section~\protect\ref{sec:qft}: these
are Ising order parameters dual to the $Z_2$ gauge theory. The
phase on the right is a conventional Fermi liquid; that on the
left is a Fermi liquid with local stripe correlations with random
phases and spin orientations, and a `topological rigidity'
associated with the suppression of stripe dislocations. In the
application to the cuprates, the horizontal axis represents
increasing doping and the critical point is near optimal doping.
States with long range spin or charge density wave order do appear
at smaller doping but these have not been shown. The phase
transition at $K=K_c$ should be distinguished from the
conventional SDW ordering transition in a superconductor; this
occurs at lower doping and was used recently by one of us
\protect\cite{dsz,zds} to predict and explain the results of
neutron scattering and STM experiments.} \label{phase}
\end{figure}

We reiterate that there is no long range spin or charge density
wave order on either side of this fractionalization transition,
and it is entirely associated with the suppression of the defects
in Fig~\ref{dislocation}; see Fig~\ref{phase}.
At dopings lower than those shown in Fig~\ref{phase}, the cuprates
possess phases with a variety of possible long range orders ({\em
e.g.\/} SDW or CDW) which were reviewed in Ref.~\onlinecite{zds}.
These transitions are associated with conventional order
parameters, and we will not consider them in this paper, as
theories are already available in the literature.

\section{Quantum field theories} \label{sec:qft}

We are interested in describing the universal properties of the
transition at $K=K_c$. These turn out to be controlled by two
distinct characteristics of the electronic system.
\newline ({\em i\/}) Tetragonal or orthorhombic crystal symmetry:
With tetragonal symmetry we have to simultaneously consider the
separate $Z_2$ gauge theories associated with fractionalization
transitions in both $\Phi_{x \alpha}$ and $\Phi_{y \alpha}$. With
orthorhombic symmetry, only one of these will be selected (the
``stripes'' have a preferred direction).
\newline ({\em ii\/}) Presence or absence of long-range Coulomb
interactions: As we argue below, an important coupling of the
critical $Z_2$ gauge degrees of freedom to the fermions arises
from long-wavelength deformations of the shape of the Fermi
surface. As is well known, in the presence of long-range Coulomb
interactions the ``dilatational'' mode, involving local changes in
the area of the Fermi surface, is strongly suppressed. For
completeness we will also consider the case of fermions with only
short-range interactions (perhaps the Coulomb interactions have
been screened by a metallic gate), where the coupling to the
dilatational mode has to be included.

With these two criteria, four distinct categories emerge, which we
label A, B, C, D, as summarized in Table~\ref{table}.
\begin{table}
\caption{\label{table} Classification of quantum field theories.}
\begin{ruledtabular}
\begin{tabular}{c|c|c|c}
Case & Conditions & Fields & Action \\
\hline
A & Orthorhomic symmetry and & $\phi_x$ & $\mathcal{S}_x$ \\
~ & long-range Coulomb interactions & ~ & ~ \\
\hline B & Tetragonal symmetry and & $\phi_x$, $\phi_y$, $\psi_I$
& $\mathcal{S}_{xy} +
\mathcal{S}_I $ \\
~ & long-range Coulomb interactions & ~ & ~ \\
\hline C & Orthorhombic symmetry and & $\phi_x$, $\psi$ &
$\mathcal{S}_{x} +
\mathcal{S}_{\psi} $ \\
~ & short-range interactions & ~ & ~ \\
\hline D & Tetragonal symmetry and & $\phi_x$, $\phi_y$, &
$\mathcal{S}_{xy} +
\mathcal{S}_I $ \\
~ & short-range interactions & $\psi_I$, $\psi$ & $+ \mathcal{S}_{\psi}^{\prime}$ \\
\end{tabular}
\end{ruledtabular}
\end{table}
We will describe these cases in the subsections below.

\subsection{Case A}
\label{a} We begin with case A, for which the critical theory has
the simplest form. As we noted in Section~\ref{sec:op}, the
transition at $K=K_c$ is described by a $Z_2$ gauge theory. We
know that the pure 2+1 dimensional $Z_2$ gauge theory is dual a
2+1 dimensional Ising model\cite{wegner,fradkin,Senthil99}; we
represent this Ising order parameter by a real scalar field
$\phi_x$. For now, we neglect the independent $Z_2$ gauge theory
associated with $\Phi_{y \alpha}$, because orthorhombic symmetry
has oriented all the spin and charge density waves along $\Phi_{x
\alpha}$. The action for $\phi_{x}$ in the vicinity of the
transition is the familiar field theory for the Ising critical
point:
\begin{equation}
\mathcal{S}_{x} = \int d^2 r d \tau \left[ \frac{1}{2} \left(
\partial_{\tau} \phi_x \right)^2 + \frac{1}{2} \left( \nabla_r \phi_x \right)^2 +
\frac{s}{2}  \phi_x^2 + \frac{u}{24} \phi_x^4 \right]; \label{sx}
\end{equation}
we have chosen the spatial units so that the velocity of $\phi_x$
excitations is unity in both directions, $u$ is a quartic
non-linearity, and the quadratic coupling $s \sim K - K_c$ is the
tuning parameter across the transition. The field $\phi_x$ can be
viewed as the creation/annihilation operator for the dislocations:
$\phi_x$ is condensed in the confining phase where the
dislocations proliferate, and $\phi_x$ has short-range
correlations in the deconfined phase where dislocations are
suppressed.

In principle, we also need to account for the ``matter'' fields
$\theta_x$, $n_{x\alpha}$ in (\ref{z2flux}) near this confinement
transition. Closely related $Z_2$ gauge models have been
considered in the literature\cite{fradkin,Lammert93,Senthil99},
and it has been established that their fluctuations do not change
the universality class of the transition. Their presence does
modify the effective couplings in (\ref{sx}), but the confinement
transition continues to be described by the Ising field theory
(\ref{sx}) all along the phase boundary.

The main question we need to address here is the coupling between
$\phi$ and the electronic excitations near the Fermi surface; the
gapless excitations near the Fermi surface have the potential of
inducing long-range interactions which are more destructive than
the matter fields in (\ref{z2flux}). The electrons {\em do not\/}
carry any Ising gauge charges, and couple to degrees of freedom of
interest here only via the bilinears in (\ref{e1}) and (\ref{e2}):
\begin{equation}
-\lambda_1 \sum_{\bf r} c^{\dagger}_\varsigma ({\bf r})
\tau^{\alpha}_{\varsigma \varsigma'} c_{\varsigma'} ({\bf r})
S_{\alpha} ({\bf r}) - \lambda_2 \sum_{\bf r}
c^{\dagger}_\varsigma ({\bf r})  c_\varsigma ({\bf r}) Q_{{\bf a}}
({\bf r}),
\end{equation}
where $\tau^{\alpha}$ are the spin Pauli matrices. We now imagine
integrating out $\Phi_{x \alpha}$ and generating terms which
couple the electrons to the $Z_2$ gauge degrees of freedom. The
electrons are gauge singlets, and so can only couple to the
physical $Z_2$ gauge flux: the simplest such coupling is one in
which the coupling $K$ in (\ref{z2flux}) becomes ${\bf r},\tau$
dependent via its dependence on the local fermion bilinears
\begin{equation}
(K,s) \rightarrow (K,s) + \lambda_3 c_{\varsigma}^{\dagger} ({\bf
r}) c_\varsigma ({\bf r}) + \lambda_4 c_{\varsigma}^{\dagger}
({\bf r}) \nabla_{\bf r}^2 c_\varsigma ({\bf r}) + \ldots,
\label{kmap}
\end{equation}
where the ellipses represent terms with higher derivatives; we
have to include all such terms because the typical electron
momentum is of order the Fermi momentum, and this is not small.
Upon performing the duality transition to the Ising field
$\phi_x$, the mapping applies to $s$, the co-efficient of
$\phi_x^2$, via its dependence upon $K$, and this is also noted in
(\ref{kmap}). We can now integrate over the fermionic degrees of
freedom and examine the structure of the terms which modify
$\mathcal{S}_x$: it is not difficult to see that apart from the
modification of the numerical values of the couplings in
(\ref{sx}), all such terms are formally irrelevant. The leading
non-analytic terms induced by the low energy fermionic excitations
arise from long wavelength deformations of the Fermi surface: in
the presence of long-range Coulomb interactions, all such
excitations allowed by symmetry to couple to $\phi_x^2$ are
suppressed. So we reach the important conclusion that the critical
theory for case A is specified by (\ref{sx}).

There is one important caveat to this conclusion that deserves
highlighting. Our analysis has neglected topological Berry phase
terms that could be present in the coupling between $\phi_x$, the
fermions, and the $Z_2$ gauge field $\sigma_{ij}$. Similar phase
factors do arise in other analyses of $Z_2$ gauge theories
\cite{Senthil99}. We leave the study of this issue to future work,
and restrict our attention here to couplings associated with
(\ref{kmap}).

\subsection{Case B}
\label{b}
The higher tetragonal symmetry has two important
consequences: we have two consider fractionalization in both
$\Phi_{x \alpha}$ and $\Phi_{y \alpha}$ simultaneously, and
distortions of the Fermi surface with $B_1$ symmetry do couple to
the critical degrees of the freedom even in the presence of
long-range Coulomb interactions.

There are now two dual Ising fields, $\phi_x$ and $\phi_y$,
associated with the independent $Z_2$ gauge fields in the $x$ and
$y$ directions; simple symmetry considerations show that
$\mathcal{S}_x$ in (\ref{sx}) is replaced by
\begin{eqnarray}
\mathcal{S}_{xy} &=& \int d^2 r d \tau \Biggl[\frac{1}{2} \left(
\partial_{\tau} \phi_x \right)^2 + \frac{1}{2} \left(
\partial_{\tau} \phi_y \right)^2 + \frac{1}{2} \left( \nabla_r \phi_x \right)^2
\nonumber \\ &~&~~~~~~~~~+ \frac{1}{2} \left( \nabla_r \phi_y
\right)^2 + \frac{s}{2}  \left( \phi_x^2+ \phi_y^2 \right)
\nonumber \\ &~&~~~~~~~~~+ \frac{u}{24} \left(\phi_x^2 + \phi_y^2
\right)^2 + \frac{v}{4} \phi_x^2 \phi_y^2 \Biggr]; \label{sxy}
\end{eqnarray}
for simplicity, we have neglected a possible difference in the
velocity of the $\phi_{x}$ modes in the $x$ and $y$ directions.
The critical properties of the theory $\mathcal{S}_{xy}$ are well
understood \cite{amnon}: they are described by a stable fixed
point with $v^{\ast} = 0$ and a global O(2) (or XY) symmetry of
rotations between $\phi_{x,y}$.

The fundamental new phenomenon here is that low energy
deformations of the Fermi surface destabilize the critical point
just noted. Although long-range Coulomb interactions do suppress a
local uniform dilatational change in the Fermi surface, they do
not suppress shape deformations which preserve the local volume.
In particular, the tetragonal symmetry allows an oscillation of
the Fermi surface with $B_1$ (or $d_{x^2-y^2}$) symmetry which has
singular contributions at low energy, as has been emphasized
recently by Oganesyan {\em et al.} \cite{vadim}; moreover, by
symmetry, this deformation will couple linearly to $\phi_x^2 -
\phi_y^2$. Integrating the Fermi surface deformations out by a
standard procedure \cite{hertz,vadim}, we obtain the following
non-analytic term in the effective action:
\begin{equation}
\lambda_5 \int \frac{d^2 q}{(2 \pi)^2} \int \frac{d \omega}{2 \pi}
\frac{|\omega|}{|{\bf q}|} \left| \phi_x^2 ({\bf q}, \omega) -
\phi_y^2 ({\bf q}, \omega) \right|^2 ; \label{l5}
\end{equation}
the $|\omega|/|{\bf q}|$ factor is the standard density of states
of long-wavelength Fermi surface oscillations in dimensions $\geq
2$. A simple scaling argument now gives the scaling dimension of
the coupling $\lambda_5$ at the $v^{\ast} = 0$ fixed point of
$\mathcal{S}_{xy}$. The $|\omega|/|{\bf q}|$ factor implies that
(\ref{l5}) couples $\left(\phi_x^2 ({\bf r}, \tau) - \phi_y^2
({\bf r}, \tau) \right)$ to itself at distinct spacetime points,
which precludes any additional operator renormalization;
consequently
\begin{equation}
\mbox{dim} [ \lambda_5 ] = d+1 - 2 \mbox{dim} [ \phi_x^2 -
\phi_y^2 ]; \label{scale1}
\end{equation}
here and henceforth we will state such scaling dimensions for
general spatial dimensions $d$, although our interest is in $d=2$.
The scaling dimension of $\phi_x^2 - \phi_y^2$ can in turn be
related to the anisotropy crossover exponent \cite{amnon,cross},
$\phi_T$
\begin{equation}
\mbox{dim}[\phi_x^2 - \phi_y^2] = d+1 - \frac{\phi_T}{\nu}
\label{scale2}
\end{equation}
where $\nu$ is the correlation length exponent of the $v^{\ast}=0$
fixed point of $\mathcal{S}_{xy}$. Using the known numerical
values\cite{cross,xynu} of these exponents in $d=2$, $\phi_T =
1.184$ and $\nu = 0.67155$, we obtain $\mbox{dim} [\lambda_5 ] =
0.52 > 0$. So we have established our earlier claim that $B_1$
deformations of the Fermi surface constitute a relevant
perturbation.

It is now necessary to extend the renormalization group to include
contributions higher order in $\lambda_5$. However, a simple
one-loop analysis shows that this is not a stable procedure
\cite{amnon2}: at order $\lambda_5^2$ we generate additional
quartic terms which have a different $\omega$, ${\bf q}$
dependence than that in (\ref{l5}), and new relevant terms are
generated at each successive order. The solution to this conundrum
is to introduce a new real scalar field, $\psi_I$, as a
Hubbard-Stratonovich decoupling of (\ref{l5}): this field will be
a measure of the amplitude of $B_1$ deformations of the Fermi
surface. Note that we expect $\lambda_5 > 0$, and so to make the
Hubbard-Stratonovich transformation positive definite, it is
necessary to subtract a $(\phi_x^2 - \phi_y^2)^2$ term before
decoupling, and to compensate for this by corresponding changes to
$u$,$v$. In this manner we get the action
\begin{eqnarray}
\mathcal{S}_I &=& \frac{1}{2} \int \frac{d^2 q}{(2 \pi)^2} \int
\frac{d \omega}{2 \pi} \left( 1 + a_I \frac{|\omega|}{|{\bf q}|}
\right) \left| \psi_I ({\bf q}, \omega ) \right|^2 \nonumber \\
&~&~~~~~~~~- \frac{g}{2} \int d^2 r d \tau \psi_I \left( \phi_x^2
- \phi_y^2 \right) ; \label{si}
\end{eqnarray}
our final effective action for case B is $\mathcal{S}_{xy} +
\mathcal{S}_I$. In obtaining $\mathcal{S}_I$ we have expanded the
coefficient of $\psi_I^2$ for small $\lambda_5 \sim a_5 $, and
rescaled $\psi_I$ to make the co-efficient of the analytic
$\psi_I^2$ term unity. We now have two new coupling constants, $g$
and $a_I$, and it can be checked that $\lambda_5 > 0$ implies $a_I
> 0$. Further, at the fixed points of $\mathcal{S}_{xy}$,
we will show in Section~\ref{rgb} that $\mbox{dim}[a_I ]=
\mbox{dim}[\lambda_5]$.

The most important property of $\mathcal{S}_{xy}+\mathcal{S}_I$ is
that this action is stable under renormalization, and no new
relevant terms are generated at any order. Consequently, it can be
subjected to a standard field-theoretic renormalization group
analysis, and this will be presented in Section~\ref{sec:rg}.

\subsection{Case C}
\label{c}
The analysis of the remaining cases is a straightforward
generalization of the discussion in Section~\ref{b}.

As in Case A, the orthorhombic symmetry implies that we need only
one Ising field $\phi_x$. However, because of the absence of
long-range Coulomb interactions, dilatational changes in the local
Fermi surface volume with full lattice symmetry do carry large low
energy spectral weight. We measure these changes by the real
scalar field $\psi$, and obtain its effective action
\begin{eqnarray}
\mathcal{S}_{\psi} &=& \frac{1}{2} \int \frac{d^2 q}{(2 \pi)^2}
\int \frac{d \omega}{2 \pi} \left( 1 + a \frac{|\omega|}{|{\bf
q}|}
\right) \left| \psi ({\bf q}, \omega ) \right|^2 \nonumber \\
&~&~~~~~~~~- \frac{g}{2} \int d^2 r d \tau \psi \phi_x^2 .
\label{spsi}
\end{eqnarray}
The effective action for case C is now $\mathcal{S}_{x} +
\mathcal{S}_{\psi}$. An argument similar to that in
Section~\ref{b} can be used to compute the scaling dimension of
$a$: as in (\ref{scale1},\ref{scale2}) we find that the fixed
point describing the transition of $\mathcal{S}_x$
\begin{equation}
\mbox{dim}[a] = \frac{2}{\nu} - d -1 = \frac{\alpha}{\nu}
\label{spsia}
\end{equation}
where $\alpha$, $\nu$ are the standard exponents of the 2+1
dimensional Ising model. Using their known values \cite{isingnu},
we obtain $\mbox{dim}[a] = 0.174 > 0$. So the fixed point of case
A is unstable to a Fermi surface with short-range interactions,
and there is new universality class of critical behavior.

We note that the discussion in this subsection, with zero sound
modes of a Fermi surface coupling to the square of an order
parameter, bears a passing resemblance to a discussion of the
influence of phonon modes in the superfluid state of bosons
presented by Frey and Balents \cite{balents}

\subsection{Case D}
\label{d} With tetragonal symmetry and no long-range Coulomb
interactions, both dilatational and $B_1$ modifications of the
Fermi surface have to be included. So the full effective action is
$\mathcal{S}_{xy} + \mathcal{S}_I + \mathcal{S}_{\psi}^{\prime}$
where
\begin{eqnarray}
\mathcal{S}_{\psi}^{\prime} &=& \frac{1}{2} \int \frac{d^2 q}{(2
\pi)^2} \int \frac{d \omega}{2 \pi} \left( 1 + a
\frac{|\omega|}{|{\bf q}|}
\right) \left| \psi ({\bf q}, \omega ) \right|^2 \nonumber \\
&~&~~~~~~~~- \frac{g}{2} \int d^2 r d \tau \psi \left( \phi_x^2 +
\phi_y^2 \right) . \label{spsip}
\end{eqnarray}
At the $v^{\ast} = 0$ fixed point of $\mathcal{S}_{xy}$, we can
show, as in Section~\ref{c}, that $\mbox{dim}[a] = \alpha/\nu$,
where $\alpha$, $\nu$ are now critical exponents of the 2+1
dimensional XY model. Numerically\cite{xynu} we have
$\mbox{dim}[a] = -0.022 < 0$, and so $a$ is irrelevant at this
fixed point. However, the perturbations in $\mathcal{S}_I$ remain
relevant, and we cannot a priori conclude that the terms in
$\mathcal{S}_{\psi}^{\prime}$ will also be irrelevant at any
possible stable fixed point of $\mathcal{S}_{xy} + \mathcal{S}_I$;
so it is necessary to include (\ref{spsip}) in the analysis of
Case D.

\section{Conventional order parameters without Fermi surface
damping} \label{sec:sdw}

This section makes a brief detour to a separate class of quantum
phase transitions mentioned in Section~\ref{sec:intro}: those with
conventional order parameters which do not couple linearly to low
energy excitations on the Fermi surface. We will consider here a
simple example of such a transition, although the results
discussed here are far more general: an SDW ordering transition
with the order parameters defined as in (\ref{e1}), from a Fermi
liquid state with $\langle \Phi_{x,y\alpha} \rangle = 0$ to an SDW
state with $\langle \Phi_{x,y\alpha} \rangle \neq 0$. Further, we
assume that the values of ${\bf K}_{x,y}$ are such that they
cannot connect any two points on the Fermi surface; consequently
the $\Phi_{x,y\alpha}$ fluctuations are undamped and the quantum
critical modes are strongly coupled. The effective action for
$\Phi_{x,y\alpha}$ will have a structure similar to that of
$\mathcal{S}_x$, $\mathcal{S}_{xy}$ in (\ref{sx}), (\ref{sxy}),
apart from changes in the quartic terms which are similar to those
in (\ref{vphi}).

It should now be clear that the zero momentum composite fields
$|\Phi_{x,y\alpha}|^2$ can couple to long-wavelength deformations
of the Fermi surface in much the same way as the composite fields
$\phi_{x,y}^2$ did in Section~\ref{sec:qft}. For orthorhombic
symmetry and short-range interactions (as in Case C), only
$\Phi_{x\alpha}$ is critical (say), and its coupling to the
long-wavelength particle-hole continuum leads to the term
\begin{equation}
\lambda_6 \int \frac{d^2 q}{(2 \pi)^2} \int \frac{d \omega}{2 \pi}
\frac{|\omega|}{|{\bf q}|} \left| \Phi_{x\alpha}^2 ({\bf
q},\omega) \right|^2, \label{l6}
\end{equation}
which is the analog of (\ref{l5}). Just as in
(\ref{spsi},\ref{spsia})
\begin{equation}
\mbox{dim}[\lambda_6] = \frac{\alpha}{\nu}, \label{l6d}
\end{equation}
and so the coupling in (\ref{l6}) is relevant if and only if
$\alpha > 0$. Most conventional order parameters have $\alpha < 0$
in 2+1 dimensions, in which case the Fermi surface has no
influence on the critical theory. The main exception is the Ising
case, found {\em e.g.} at a two sublattice CDW ordering
transition, in which case the theory will identical to that in
Section~\ref{c}.

For a system with tetragonal symmetry, there is stronger coupling
to quadrupolar distortions of the Fermi surface, just as discussed
in Section~\ref{b}. The analog of (\ref{l5}) is the term
\begin{equation}
\lambda_7 \int \frac{d^2 q}{(2 \pi)^2} \int \frac{d \omega}{2 \pi}
\frac{|\omega|}{|{\bf q}|} \left| \left|\Phi_{x\alpha} \right|^2
({\bf q}, \omega) - \left|\Phi_{y\alpha}\right|^2 ({\bf q},
\omega) \right|^2 , \label{l7}
\end{equation}
and the scaling dimension of $\lambda_7$ is related to an
anisotropy crossover exponent; this always such that $\lambda_7$
is strongly relevant. The analysis of the resulting theory will be
similar to Case B above, although we will not present it
explicitly.

\section{Renormalization group analysis}
\label{sec:rg}

This section returns to the collective mode fractionalization
transitions considered in Sections~\ref{sec:op} and
~\ref{sec:qft}; closely related analyses will apply to the cases
mentioned in Section~\ref{sec:sdw}.

Case A is described by the purely bosonic theory (\ref{sx}), whose
properties are very well known: its critical point is in the
universality class of the 2+1 dimensional Ising model. So we move
directly to case B in the first subsection, and mention the
straightforward generalization to the other cases later.

\subsection{Case B}
\label{rgb} Renormalization group (RG) analyses of models with a
structure similar to $\mathcal{S}_{xy} +\mathcal{S}_I$ have
appeared elsewhere in the literature (see {\em e.g.}
Ref.~\onlinecite{balents}), and so we can omit most of the
technical details. We consider the rescaling transformation
\begin{equation}
r = r' e^{\ell} ~~~;~~~\tau=\tau' e^{z \ell}
\end{equation}
under which the fields $\phi_{x,y}$, $\psi_I$ transform as
\begin{eqnarray}
\phi_{x,y} (r,\tau) &=& e^{-(d+z-2+\eta)\ell/2}\phi_{x,y}^{\prime}
(r',\tau') \nonumber \\ \psi_I (r,\tau) &=&
e^{-(d+z+\theta)\ell/2}\psi_I^{\prime} (r',\tau');
\end{eqnarray}
here $z$, $\theta$, and $\eta$ are $\ell$-dependent parameters to
be determined by loop corrections to the effective action under
the RG transformation---the values of these parameters at any
fixed point of the RG equations will specify the corresponding
critical exponents. The one-loop graphs have the same topology as
those considered in Ref.~\onlinecite{balents}, and we defer
details on the evaluation of the loop integrals to
Appendix~\ref{oneloop}; the results are
\begin{eqnarray}
\frac{da_I}{d\ell} &=& (1-z-\theta)a_I \nonumber \\
 \frac{dg}{d\ell} &=&
\frac{(4 - d - z - \theta - 2 \eta)}{2} g - \left(\frac{u}{3} -
\frac{v}{2} \right) g \mathcal{I} (0,a_I) \nonumber \\ &+& g^3
\mathcal{I} (1,a_I) \nonumber \\
\frac{du}{d\ell} &=& (4 - d - z - 2 \eta) u - \frac{(10 u^2 + 6 u
v + 9 v^2) }{6} \mathcal{I} (0,a_I) \nonumber \\ &+& 6 u g^2
\mathcal{I} (1,a_I)- 6 g^4
\mathcal{I} (2,a_I) \nonumber \\
\frac{dv}{d\ell} &=& (4 - d - z - 2 \eta) v - \frac{( 4 u v + 3
v^2) }{2} \mathcal{I} (0,a_I) \nonumber \\ &-& \frac{2(4 u+ 3
v)}{3} g^2 \mathcal{I}
(1,a_I) \nonumber \\
\theta &=& - g^2 \mathcal{I} (0,a_I) \nonumber \\
\eta &=& g^2 \mathcal{J} (a_I) \nonumber \\
z &=& 1+ \frac{g^2 \mathcal{K} (a_I)-\eta}{2}\label{rg1}
\end{eqnarray}
where $\mathcal{I}_ (m,a_I)$ ($m$ integer), $\mathcal{J} (a_I)$,
and $\mathcal{K} (a_I)$ are scheme dependent functions which are
specified in Appendix~\ref{oneloop}. We have assumed that the
strongly relevant ``mass'' term $s$ has been tuned to its critical
value, and are examining above only the flow within this critical
manifold. The first of the equations in (\ref{rg1}) is the most
important: it is an exact equation (although the values of $z$ and
$\theta$ do have corrections at higher order) which follows from
the fact that there are no loop corrections to the non-analytic
term with coefficient $a_I$ in the effective action.

First, let us examine (\ref{rg1}) in the subspace with $a_I=0$,
where the long-range forces associated with the Fermi surface are
absent. The results in Appendix~\ref{rg1}) show that $\mathcal{I}
(m,0) = \mathcal{C}$ independent of $m$, and $\mathcal{J}(0) =
\mathcal{K}(0)=0$; here $\mathcal{C}$ is a scheme-dependent
constant whose value does not effect the physical properties. It
is then easy to search for fixed points of (\ref{rg1}): there are
8 fixed points, and all of them obey $z=1$ (this is exact to all
orders for $a_I=0$) and $\eta=0$ (to this order). The fixed points
are:
\begin{eqnarray}
&(G_1)&~u^{\ast} = 0~;~v^{\ast} = 0~;~
g^{\ast 2} = 0 \nonumber \\
&(G_2)&~u^{\ast} = \frac{3\epsilon}{\mathcal{C}}~;~v^{\ast} =
-\frac{2 \epsilon}{\mathcal{C}}~;~ g^{\ast 2} = \epsilon
\nonumber \\
&(C_1)&~u^{\ast} = \frac{\epsilon}{3\mathcal{C}}~;~v^{\ast} =
\frac{2 \epsilon}{9\mathcal{C}}~;~ g^{\ast 2} = 0
\nonumber \\
&(C_2)&~u^{\ast} = \frac{10 \epsilon}{3\mathcal{C}}~;~v^{\ast} =
-\frac{16 \epsilon}{9\mathcal{C}}~;~ g^{\ast 2} = \epsilon
\nonumber
\\ &(XY_1)&~u^{\ast} = \frac{3\epsilon}{5\mathcal{C}}~;~v^{\ast} = 0~;~
g^{\ast 2} = 0 \nonumber
\\ &(XY_2)&~u^{\ast} = \frac{12\epsilon}{5\mathcal{C}}~;~v^{\ast} = -\frac{6
\epsilon}{5\mathcal{C}}~;~ g^{\ast 2} = \frac{3
\epsilon}{5\mathcal{C}}
\nonumber \\
&(I_1)&~u^{\ast} = \frac{2\epsilon}{3\mathcal{C}}~;~v^{\ast} =
-\frac{2 \epsilon}{9\mathcal{C}}~;~ g^{\ast 2} = 0
\nonumber \\
&(I_2)&~u^{\ast} = \frac{5\epsilon}{3\mathcal{C}}~;~v^{\ast} =
-\frac{8 \epsilon}{9\mathcal{C}}~;~ g^{\ast 2} =
\frac{\epsilon}{3\mathcal{C}},
 \label{fp2}
\end{eqnarray}
where $\epsilon=3-d$. The physical meaning of these fixed points
becomes clear after integrating out the $\psi_I$ field in
$\mathcal{S}_{xy}$ at $a_I=0$: for this value of $a_I$ this yields
an effective action with the same structure as $\mathcal{S}_{xy}$
in (\ref{sxy}) but with modified coupling constants
\begin{eqnarray}
\overline{u} &=& u - 3 g^2 \nonumber \\
\overline{v} &=& v + 2 g^2. \label{ubar}
\end{eqnarray}
The reader can now observe that the fixed points in (\ref{fp2})
all separate into pairs with equal values of $\overline{u}$ and
$\overline{v}$, and our notation for the fixed points has been
chosen to make this evident: $G_1$ and $G_2$ form one such pair,
and similarly for the remainder.

At this stage, we can compare results with the analysis of
Aharony\cite{aharony,amnon} of the consequences of a cubic
anisotropy on fixed points with O($n$) symmetry: he analyzed the
analog of the model $\mathcal{S}_{xy}$ for $n$-component fields,
but with a somewhat different notation. The fixed points
(\ref{fp2}) are in complete agreement with his results: $G_{1,2}$
are the Gaussian fixed points, $C_{1,2}$ are the ``cubic'' fixed
points, $XY_{1,2}$ are the XY fixed points with O(2) symmetry and
no cubic anisotropy, and $I_{1,2}$ are the Ising fixed points at
which the fields $\phi_{1,2}$ are decoupled. As shown by Aharony,
an XY fixed point is stable for the case of a two-component field
being considered here.

Here, we have to perform some further analysis to distinguish
between the two possible $XY$ fixed points: these two have the
same values of $\overline{u}$, $\overline{v}$, but only one has
$g^{\ast} \neq 0$. A standard computation shows that fixed point
$XY_2$ with $g^{\ast} \neq 0$ is the only stable fixed point among
all those in (\ref{fp2}). The theory prefers the fixed point with
$g^{\ast} \neq 0$ because the fluctuations of $\psi_I \sim
\phi_1^2 - \phi_2^2$ are characterized by a divergent
susceptibility at the critical point \cite{hhm}.

We are now in a position to consider non-zero values of $a_I$. For
small $a_I$, the above analysis shows that near the fixed point
$XY_2$
\begin{equation}
\frac{da_I}{d\ell} = \frac{3\epsilon}{5} a_I; \label{dai}
\end{equation}
In other words, $a_I$ is a {\em relevant} perturbation with
$\mbox{dim}[a_I] = 3 \epsilon /5$. We argued in Section~\ref{b}
that $\mbox{dim}[a_I] = 2 \phi_T /\nu - d -1$, and it can be
checked that this is consistent with our present result and the
$\epsilon$ expansion result\cite{amnon} for $\phi_T/\nu$.

\begin{figure}
\centerline{\includegraphics[width=3in]{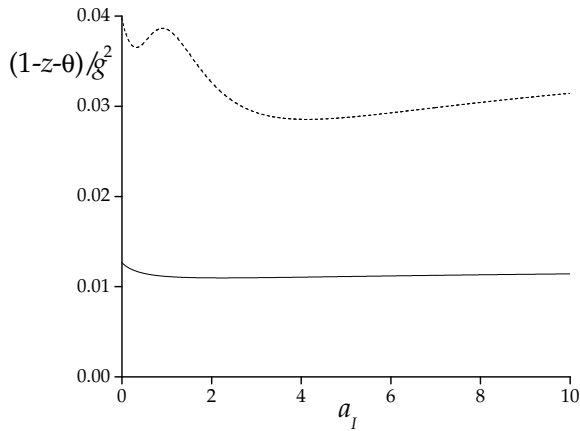}} \caption{Plot
of $(1-z-\theta)/g^2$ as determined by (\protect\ref{rg1}). There
is no zero crossing, and hence no finite, positive, non-zero $a_I$
fixed point. The full line is the minimal subtraction scheme,
while the dashed line is the fixed dimension scheme.}
\label{plotai}
\end{figure}

For general $a_I$, we have to consider the complete flow equations
in (\ref{rg1}). While these are quite complicated, it is clear
from the first flow equation that a finite and non-zero $a_I$
fixed point, $a_I=a_I^{\ast}$ can only exist if and only if there
is a solution to $1-z-\theta=0$. From (\ref{rg1}) we observe that
at the one loop order, this condition translates into an equation
for $a_I^{\ast}$ alone:
\begin{equation}
2 \mathcal{I}(0, a_I^{\ast}) + \mathcal{J} (a_I^{\ast}) -
\mathcal{K}(a_I^{\ast}) = 0. \label{eqai}
\end{equation}
Using the results in Appendix~\ref{oneloop}, we plot half the left
hand side of this equation in Fig~\ref{plotai} for two different
renormalization schemes.
No solution is found in either case, or in the additional cutoff
schemes discussed in Appendix~\ref{oneloop}. The value of
$(1-z-\theta)$ does initially decrease from its value at $a_I=0$
which led to (\ref{dai}), but this decrease is not strong enough
to lead to a zero crossing. So we have not found a finite $a_I$
fixed point, but we cannot rule out the possibility that one will
appear upon including terms higher order in $g$, $u$, $v$.

It is clear now that any non-zero, positive $a_I$ runs away to
$a_I = \infty$, and we have to consider the theory in this limit.
An examination of the limiting values of the results in
Appendix~\ref{oneloop} showed that there were no finite fixed
point values for $u$, $v$ in the limit of large $a_I$ either. In
principle, it was possible that both $a_{I}$ and $g$ became large
in a manner that permitted $u$ and $v$ to approach a finite fixed
point value. However simple analytic and numerical analyses of
(\ref{rg1}) showed that this was not the case, and we always had
flows to strong coupling for $u$ and $v$ also. Starting close to
the stable fixed point at $a=0$, we found that the flows sent
$v\rightarrow \infty$ and $u \rightarrow -\infty$. This is
possibly suggestive of a first-order transition, but no definitive
conclusion can be reached because the RG equations themselves
breakdown once $u$ or $v$ become of order unity.

\subsection{Case C}
\label{rgc} The analysis and results for Case C are similar to
those for Case B above, and so we will be brief.

\begin{figure}
\centerline{\includegraphics[width=3in]{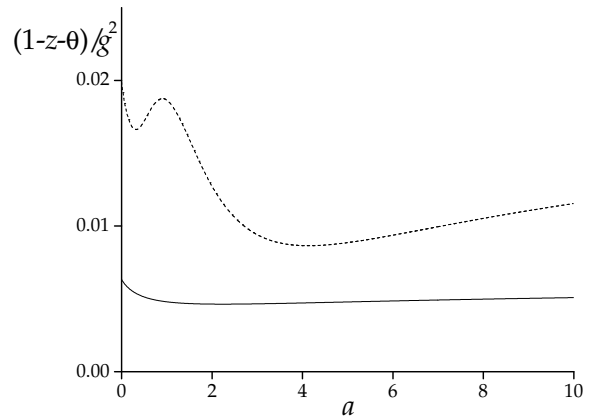}} \caption{An in
Fig~\protect\ref{plotai}, but for (\protect\ref{eqa}).}
\label{plota}
\end{figure}

The RG flow equations (\ref{rg1}) are replaced by
\begin{eqnarray}
\frac{da}{d\ell} &=& (1-z-\theta)a \nonumber \\
 \frac{dg}{d\ell} &=&
\frac{(4 - d - z - \theta - 2 \eta)}{2} g - \frac{ug}{2}
\mathcal{I} (0,a) \nonumber \\ &+& g^3
\mathcal{I} (1,a) \nonumber \\
\frac{du}{d\ell} &=& (4 - d - z - 2 \eta) u - \frac{3 u^2}{2}
\mathcal{I} (0,a) \nonumber \\ &+& 6 u g^2 \mathcal{I} (1,a)- 6
g^4
\mathcal{I} (2,a) \nonumber \\
\theta &=& - \frac{g^2}{2} \mathcal{I} (0,a) \nonumber \\
\eta &=& g^2 \mathcal{J} (a) \nonumber \\
z &=& 1+ \frac{g^2 \mathcal{K} (a)-\eta}{2},\label{rg3}
\end{eqnarray}
where the functions $\mathcal{I}$, $\mathcal{J}$, $\mathcal{K}$
remain the ones specified in Appendix~\ref{oneloop}.  These
equations do have fixed points at $a=0$, which have been specified
completely in earlier work\cite{hhm}. However, the stable fixed
point for $a=0$ is unstable to a non-zero $a$, and the condition
for a finite and non-zero $a$ fixed point is (replacing
(\ref{eqai})):
\begin{equation}
\mathcal{I}(0, a^{\ast}) + \mathcal{J} (a^{\ast}) -
\mathcal{K}(a^{\ast}) = 0. \label{eqa}
\end{equation}
The left hand side of this equation is plotted in
Fig.~\ref{plota}, and again there is no solution.
We find instead flows to strong coupling, as in Case B.

Finally, we note that the flow equations and results for Case D
are a combination of those presented already for Cases C and D,
and we will not present them explicitly.

\section{Physical Properties}
\label{sec:phys}

This section considers some simple observable physical properties
of the fractionalization transition described in
Section~\ref{sec:qft}. We will find that the properties of many
observables are quite similar to another case with an order
parameter which is a composite of the electron spin operator: the
onset of spin glass order. To help make this comparison, we review
the physical properties of the quantum spin glass transition in
Appendix~\ref{app:qsg}.

Our analysis of collective mode fractionalization theories has
found only one stable fixed point: that for case $A$ with
orthorhombic symmetry and long-range Coulomb interactions. The
other cases exhibited flows to strong coupling. This section will
study the general scaling properties of various physical
observables for case A, and also extrapolate the results to
possible strongly coupling fixed points for the other cases.

\subsection{Dynamic spin susceptibility}
\label{sec:sus}

We focus on the spin susceptibility near the wavevectors ${\bf
K}_{x,y}$; other susceptibilities of conventional order parameters
can be treated in a similar way. The order parameters associated
with these susceptibilities, $\Phi_{x,y\alpha}$, are controlled by
an effective action of the form ($\omega_n$ is a Matsubara
frequency at a temperature $T$)
\begin{equation}
\mathcal{S}_{\Phi} = \frac{T}{2} \sum_{\omega_n} \int \frac{d^2
q}{(2 \pi)^2} \left( \Gamma |\omega_n| + A {\bf q}^2 + B \right)
\left|\Phi_{x\alpha} ({\bf q}, \omega_n ) \right|^2,
\end{equation}
along with a similar term with $(x \rightarrow y)$. Unlike the
situation in Section~\ref{sec:sdw}, we are considering the case
where ${\bf K}_{x,y}$ do connect two points on the Fermi surface,
and hence there is over-damping of $\Phi_{x,y\alpha}$ fluctuations
represented by the term proportional to $\Gamma$. Also, the
$\Phi_{x,y\alpha}$ fluctuations remain non-critical with $\langle
\Phi_{x,y \alpha} \rangle = 0$ at the fractionalization
transition, and hence the parameter $B>0$.

We now have to consider the coupling of $\Phi_{x,y\alpha}$ to the
critical modes $\phi_{x,y}$. As noted at the end of
Section~\ref{a}, we ignore possible non-local topological terms,
and focus on the simplest local couplings allowed by symmetry; for
the cases with tetragonal symmetry (cases B and D), these are
\begin{eqnarray}
&& \int d^2 r d \tau \left[ \zeta_1 \left( \left|\Phi_{x
\alpha}\right|^2 + \left|\Phi_{y \alpha}\right|^2 \right) \left(
\phi_x^2 + \phi_y^2 \right) \right. \nonumber \\
&&~~~~~~~~~~+\zeta_2 \left( \left|\Phi_{x \alpha}\right|^2 -
\left|\Phi_{y \alpha}\right|^2 \right) \left( \phi_x^2 - \phi_y^2
\right), \label{zeta}
\end{eqnarray}
while for orthorhombic symmetry we merely drop the non-critical
$\phi_y$ field above. We will consider the influence of the
critical $\phi_{x,y}$ modes on $\Phi_{x,y\alpha}$ in a
perturbation theory in $\zeta_1$. For case A, where a stable fixed
point was found, such a procedure appears reasonably safe. The
validity for the other cases, with flows to strong coupling, has a
greater uncertainty and this should be kept in mind below.

The influence of the critical fluctuations on the non-critical
$\Phi_{x,y\alpha}$ susceptibility will be determined by two (and
higher) point correlators of the operators $\phi_x^2 - \phi_y^2$
and $\phi_x^2 + \phi_y^2$ for tetragonal symmetry, and $\phi_x^2$
alone for orthorhombic symmetry. Let us represent these operators
generically by $\mathcal{O}$ and its scaling dimension by
\begin{equation}
\mbox{dim}[\mathcal{O}] = d_{\mathcal{O}}. \label{DO}
\end{equation}
For case A, we have only $\mathcal{O} = \phi_x^2$ and
$d_{\mathcal{O}} = d+1-1/\nu$, where $\nu$ is the exponent of 2+1
dimensional Ising model; for other cases, the value of
$d_\mathcal{O}$ will be determined by the hypothetical
strong-coupling fixed point. In perturbation theory in
$\zeta_{1,2}$ we can write for the dynamic spin susceptibility
\begin{equation}
\chi_{\Phi} ({\bf q}, \omega) = \frac{1}{-i \Gamma \omega + A {\bf
q}^2 + B + M ({\bf q}, \omega)} \label{D1}
\end{equation}
with the ``self energy''
\begin{equation}
M({\bf q}, \omega_n) = \zeta^2 T \sum_{\epsilon_n} \int \frac{d^2
p}{(2 \pi)^2} D_\mathcal{O} ({\bf p}, \epsilon_n) \chi_{\Phi}
({\bf p} + {\bf q}, \epsilon_n + \omega_n) \label{D2}
\end{equation}
where $\zeta$ is related to $\zeta_{1,2}$, and $D_{\mathcal{O}}$
is the two-point correlator of $\mathcal{O}$. We are interested in
the behavior of $M$ at small ${\bf q}$ and $\omega$; after
inserting (\ref{D1}) into (\ref{D2}), we use the fact that $B>0$
to expand the momentum and frequency dependence of $\chi_{\Phi}$
in powers of $1/B$. For the ${\bf q}$ dependence of $M$, a simple
power-counting analysis shows that all terms in this series are
free of infrared divergences as long as $d_{\mathcal{O}} > 0$, in
which case the ${\bf q}$ dependence of $M$ is analytic and free of
critical singularities. Case A easily satisfies $d_{\mathcal{O}} >
0$, and we expect this is also true for the other cases, because a
negative $d_{\mathcal{O}}$ would imply that correlations of
$\mathcal{O}$ increase with increasing spacetime distance.

In contrast to its ${\bf q}$ dependence, the $\omega$ dependence
of $M$ is singular even for $d_{\mathcal{O}} > 0$. This is most
easily computed by analytically continuing (\ref{D2}) to real
frequencies and taking its imaginary part. A standard analysis
then shows that $M({\bf q}, \omega)$ has a ${\bf q}$ {\em
independent} singular contribution in the quantum critical region
which obeys the scaling form
\begin{equation}
M({\bf q}, \omega) \sim T^{(2 d_{\mathcal{O}}+z)/z} F_1(\omega/T)
\label{D3}
\end{equation}
(we are using $\hbar = k_B = 1$), where $F_1$ is a universal
scaling function determined by the non-trivial interacting fixed
point of the strongly-coupled quantum field theories considered in
Section~\ref{sec:qft}; we expect $\mbox{Im} F_1(y \rightarrow 0 )
\sim y$ and $F_1( y \rightarrow \infty ) \sim \mbox{sgn}(y)
|y|^{(2 d_{\mathcal{O}}+1)/z}$. Specifically, $F_1$ is related to
the local correlator of $\mathcal{O}$, which has a singular
contribution obeying the scaling form
\begin{equation}
\int \frac{d^2 p}{(2 \pi)^2} D_{\mathcal{O}} ({\bf p}, \omega_n)
\sim T^{2d_{\mathcal{O}}/z} F_2(\omega_n /T), \label{D4}
\end{equation}
(the scaling function $F_2$ is a property of the
strongly-interacting fixed point of the quantum field theories of
Section~\ref{sec:qft}) by the integral relation
\begin{eqnarray}
&~&\mbox{Im}\left[F_1(y)\right] \nonumber \\ &=& \int_{0}^{y}
\frac{d\Omega}{\pi} \Omega \mbox{Im} \left[ F_2(y - \Omega)
\right] \left( 1 +
n(\Omega) + n(y - \Omega) \right) \nonumber \\
&+& \int_{0}^{\infty} \frac{d\Omega}{\pi} \Omega \mbox{Im} \left[
F_2(\Omega+y) \right] \left( n(\Omega) - n(\Omega+y) \right)
\nonumber \\&+& \int_{y}^{\infty} \frac{d\Omega}{\pi} \Omega
\mbox{Im} \left[ F_2(\Omega-y) \right] \left(n(\Omega-y) - n(
\Omega) \right) , \label{D5}
\end{eqnarray}
where $n(\Omega) = 1/(e^{\Omega} - 1)$ is the Bose function at
unit temperature. Note that the leading factor of $\Omega$ in
(\ref{D5}) arises from the linearly density of states of low
frequency excitations implied by (\ref{D1}), upon ignoring the
influence of $M$ in $\chi_{\Phi}$ on the right hand side of
(\ref{D2}). This is permissible as long as the contribution of $M$
here is subdominant: from (\ref{D3}) we see that this is so as
long as $d_{\mathcal{O}} > 0$. If this inequality is violated,
then the perturbative expansion in $\zeta$ fails, and a self
consistent approach appears necessary.

It is interesting to note that the results (\ref{D1},\ref{D3})
imply a structure for the dynamic spin susceptibility which is
closely related to phenomenological forms that have been proposed
recently by Schr\"oder {\em et al.} \cite{almut} to describe
neutron scattering measurements on CeCu$_{6-x}$Au$_x$. Elegant
theoretical models of a ``local quantum criticality'' based on the
self-consistent treatment of quantum impurity (`Kondo') models
have also been advanced \cite{varma,si1,si2} and discussed
\cite{piers,bgg,zarand,marco} as explanations for the observations
of Schr\"oder {\em et al.}. Some important differences between the
theories of Si {\em et al.} \cite{si1,si2}  and our work here
should be noted: underlying our results is a bona fide $d+1$
dimensional quantum field theory (presented in
Section~\ref{sec:qft}), and not a quantum impurity model. Our
theory leads to a ${\bf q}$-independent singular contribution to
the dynamic spin susceptibility because $B>0$, as shown above. The
condition $B>0$ also implies the static spin susceptibility is not
divergent, in contrast to impurity model analyses in which it does
diverge at the impurity critical point \cite{si1,si2}. An
associated fact is that the exponent in (\ref{D3}) is likely to
satisfy $(2 d_{\mathcal{O}}+z)/z
> 1$, although this need not be true for the spin glass transitions
discussed in Appendix~\ref{app:qsg}. It has also been noted
\cite{bgg,marco,rg} that the divergence of the local
susceptibility could preempt a local treatment of the impurity
model critical points \cite{si1,si2}.

It should be clear that the above mechanism for `local' quantum
criticality in the dynamic spin susceptibility is quite a general
property of a situation in which the primary order parameter,
which exhibits strongly coupled bulk criticality, is coupled to
the spin operator only with non-linear terms like those in
(\ref{zeta}). As an illustration we recall in
Appendix~\ref{app:qsg} that related considerations apply to the
quantum spin glass transition.

\subsection{Fermion self energy}
\label{sec:self}

We follow the same perturbative approach discussed in
Section~\ref{sec:sus}. Singular fluctuations of $\mathcal{O}$ will
couple to the electrons via couplings like
\begin{equation}
\zeta_3 \int d^2 r d \tau \mathcal{O} ({\bf r}, \tau)
c^{\dagger}_{\varsigma} ({\bf r}, \tau) c_{\varsigma} ({\bf r},
\tau) + \ldots; \label{fse1}
\end{equation}
the ellipses represent additional terms with gradients of the
electron operators (these are not small because the typical
electron momentum resides on the Fermi surface).

We should note at the outset that the status of a perturbative
computation of the electron self energy, $\Sigma$, remains open.
For the case of overdamped fermion critical points below their
upper critical dimension, the corresponding issue was addressed by
Altshuler {\em et al.} \cite{ioffe} and Castellani {\em et al.}
\cite{sing} who concluded that the leading perturbative correction
also gave the ultimate critical singularity. The extension of this
result to the strongly coupled
 critical points being discussed here is less clear. Indeed,
 for the case of a
strongly coupled spin glass quantum critical point, the
perturbative computation of the electron self energy is known to
break down \cite{gp} (see also Appendix~\ref{app:qsg}).

With this caution in mind, we proceed with a computation of the
consequences of (\ref{fse1}). The electron self energy, $\Sigma$
to order $\zeta_3^2$ is
\begin{eqnarray}
&& \Sigma ({\bf q}, \omega_n ) = \zeta_3^2 T \sum_{\epsilon_n}
\int \frac{d^2 p}{(2 \pi)^2} D_\mathcal{O} ({\bf p}, \epsilon_n)
\nonumber \\
&&~~ \times \frac{1}{i (\epsilon_n + \omega_n) -\varepsilon({\bf
p} + {\bf q}) - \Sigma({\bf p} + {\bf q}, \epsilon_n +
\omega_n)},~~~~~~~~\label{fse2}
\end{eqnarray}
where $\varepsilon ({\bf p})$ is the electron dispersion. This
expression can be evaluated using the methods discussed in {\em
e.g.} Ref.~\onlinecite{ioffe}. The result depends upon the scaling
function associated with $D_{\mathcal{O}}$ for which we write
\begin{equation}
D_\mathcal{O} ({\bf q}, \omega) \sim T^{(2 d_{\mathcal{O}}
-d-z)/z} F_3 ( {\omega}/{T} , {{\bf q}}/{T^{1/z}} ), \label{fse3}
\end{equation}
with $F_3$ a non-trivial universal scaling function determined by
the strongly interacting fixed points of Section~\ref{sec:qft} .
As in Ref.~\onlinecite{ioffe}, the singular part of the self
energy is momentum independent at this order, as long as $z>1$.
Its frequency and temperature dependence is specified by the
scaling form
\begin{equation}
\Sigma({\bf q}, \omega) \sim T^{(2 d_{\mathcal{O}} - 1)/z} F_4 (
\omega/T), \label{fse4}
\end{equation}
with the scaling function $F_4$ given by
\begin{eqnarray}
\mbox{Im} \left[F_4 (y)\right] &=& \int_0^{\infty} \frac{d
\Omega}{\pi} \int_0^{\infty} k^{d-2} dk \mbox{Im} \left[ F_3
(\Omega,k) \right] \nonumber
\\&\times& \left( 2 n(\Omega) + f(\Omega-y) + f(\Omega+y)
\right),~~~~ \label{fse5}
\end{eqnarray}
where $n(\Omega)$ was defined below (\ref{D5}) and $f(\Omega) =
1/(e^{\Omega} + 1)$ is the Fermi function at unit temperature.

\subsection{Transport}
\label{sec:trans}

We identify two distinct contributions to the critical electrical
transport properties: the electronic quasiparticle and the bosonic
collective modes. However, the cautions about a perturbative
treatment of the coupling between these modes mentioned in
Section~\ref{sec:self} apply also to our discussion here.

The electronic contribution is characterized by the transport
relaxation time $\tau_{\rm tr}$. The electron scattering discussed
in Section~\ref{sec:self} all occurs at small momenta ${\bf p}$
and, as is well known, the transport rate then acquires an
additional factor of ${\bf p}^2$ over the quasiparticle scattering
rate in the integral over scattered momenta. Extending the scaling
arguments of Section~\ref{sec:self} we see easily that
\begin{equation}
\frac{1}{\tau_{\rm tr}} \sim T^{(2 d_{\mathcal{O}} + 1)/z}
\end{equation}

The collective mode contribution is the analog of the
``Aslamazov-Larkin'' term in the theory of fluctuating
superconductivity: fluctuations of the order parameters
$\phi_{x,y}$ may themselves contribute to the electrical current
${\bf J}$. The relationship between the dual fields $\phi_{x,y}$
and the microscopic degrees of freedom is extremely complex, and
so we will restrict ourselves here to deducing the expressions for
the current operator using symmetry arguments alone. The current
${\bf J}$ is odd under time reversal and spatial inversion, while
the fields $\phi_{x,y}$ are even under both. Furthermore, the
current should be invariant under the dual Ising symmetries
involving the change in sign of $\phi_{x}$ and/or $\phi_{y}$. This
indicates that the current operator has contributions from
operators like
\begin{equation}
J_x \sim \partial_{\tau} \phi_x \partial_x \phi_x~,~\phi_x
\partial_{\tau} \partial_x \phi_x~,~\partial_{\tau} \phi_y \partial_x
\phi_y\ldots
\end{equation}
and similarly for $J_y$. The scaling dimensions of these operators
can be determined by standard methods for case A (some combination
is related to components of the stress-energy tensor with scaling
dimension $d+1$, but the others require a $(3-d)$ expansion),
while for the other cases the scaling dimensions remain unknown
quantities to be determined by the hypothetical strong-coupling
fixed point. If we generically denote the scaling dimension of
these operators by $d_J$, the Kubo formula implies that they will
yield a singular contribution to the conductivity, $\sigma$, which
scales as
\begin{equation}
\delta\sigma \sim T^{(2d_J-d-2z)/z}
\end{equation}
Note that because the current is carried by both the electronic
and bosonic degrees of freedom, we expect that we do not have to
worry about constraints imposed by overall current conservation,
as is the case in theories of the superfluid-insulator transition
in bosonic systems\cite{book}.

\section{Conclusions}
\label{sec:conc}

This paper has described strongly interacting quantum phase
transitions in the presence of a Fermi surface. The primary
examples considered involve the phase transition described in
Fig~\ref{phase}, from a conventional Fermi liquid state in which
the dislocation in Fig~\ref{dislocation} proliferates, to a
`topologically ordered' Fermi liquid state in which these
dislocations have been suppressed. Note that there is no
long-range SDW/CDW order in either state, although the amplitude
of these orders must be locally large to allow identification of
the dislocations. The conventional SDW/CDW orders are suppressed
in both phases by `spin-wave' fluctuations and by the
proliferation of other defects, such as the hedgehogs in $n_{x
\alpha}$ and integer vortices in $\theta_x$. We discussed some
simple observable properties of such a topological phase
transition in a Fermi liquid, and found results which were quite
similar to those near quantum spin glass transitions, and to the
phenomenological `local' scaling forms proposed by Schr\"oder {\em
et al.} \cite{almut}. Despite their `local' nature, the critical
singularities are controlled by a bulk quantum field theory, and
not a quantum impurity model.

The dislocation proliferation transition discussed here has been
proposed as a candidate for a quantum critical point near optimal
doping in the cuprate superconductors \cite{zaanen,zds}. In this
context, it is interesting to note that recent scanning tunnelling
microscopy images \cite{aharon} of the surface of
Bi$_2$Sr$_2$CaCu$_2$O$_{8+\delta}$ clearly show dislocations with
the geometry of Fig~\ref{dislocation} (see Fig 2c of
Ref.~\onlinecite{aharon}).

The main shortcoming of our analysis here has been runaway flow of
the renormalization group flow to a non-perturbative regime for
the interesting cases. It is possible that this signals a more
fundamental change in the nature of excitations  near the Fermi
surface than is accounted for by our formalism. We integrate out
the low energy fermions at an early stage, and the strong coupling
situation may benefit from an analysis in which quasiparticle
degrees of freedom are retained more explicitly. Future work in
this direction, along the lines of Refs.~\onlinecite{metzner},
appears desirable.

\begin{acknowledgments} We thank A.~Aharony, A.~Chubukov, E.~Demler,
M.~P.~A.~Fisher, M.~Grilli, W.~Metzner, A.~Millis, V.~Oganesyan,
T.~Senthil, Q.~Si, M.~Vojta, and J.~Zaanen for useful discussions.
This research was supported by US NSF Grant DMR 0098226. TM was
supported by a Grant-in-Aid from the Ministry of Education,
Culture, Sports, Science and Technology of Japan.

\end{acknowledgments}

\appendix

\section{Quantum spin glass transition}
\label{app:qsg}

The quantum spin glass transition in metallic systems has been
studied in a number of recent works
\cite{sy2,sro,sg,gp,gps,olivier,marcello,denis}. Here we review
some physical properties whose scaling structure resembles the
results in Section~\ref{sec:phys}.

The central actor in the theory of the spin glass transition is
the order parameter functional
\begin{equation}
Q_{\alpha\beta}^{ab} ({\bf r}, \tau_1 , \tau_2) \sim
 S_{\alpha}^a ({\bf r}, \tau_1) S_{\beta}^b ({\bf r}, \tau_2)
  \label{qab}
\end{equation}
where $a,b$ are replica indices. In the quantum critical region,
this functional has an expectation value of the form
\begin{equation}
\langle Q_{\alpha\beta}^{ab} ({\bf r}, \tau_1 , \tau_2) \rangle =
\delta_{\alpha\beta} \delta^{ab} D (\tau_1 - \tau_2 ) \label{qab2}
\end{equation}
If the quantum critical point obeys hyperscaling properties, then
Fourier transform of $D(\tau)$ will obey ``$\omega/T$ scaling'':
\begin{equation}
D(\omega) = T^{\mu} F_5 (\omega/T) \label{qab3}
\end{equation}
It should be noted that the mean-field solution of the metallic
spin glass does not obey (\ref{qab3}), and instead has a
characteristic frequency scale $\sim T^{3/2}$. However, other
mean-field spin glass models which account for proximity to a Mott
insulator do obey $\omega/T$ scaling\cite{sy2,gp,gps,olivier};
These latter solutions have the value $\mu=0$, and $D(\omega)$ has
the form
\begin{eqnarray}
D (\omega) &\propto& \ln \left(\frac{\Lambda}{|\omega|} \right) -
\int_0^{\infty}  d\Omega \mathcal{P} \left( \frac{2 \Omega
f(\Omega) }{\Omega^2 - (\omega/T)^2} \right)\nonumber
\\&~&~~~~~ + \frac{i\pi}{2} \tanh\left(\frac{\omega}{2T} \right),
\label{qab4}
\end{eqnarray}
where $f(\Omega)$ was defined below (\ref{fse5})---for this value
of $\mu$ there is an additive real contribution dependent upon the
cutoff $\Lambda$, but the remainder is a universal function of
$\omega/T$, and the function $D(\omega)$ is analytic as $\omega
\rightarrow 0$ at any nonzero $T$. We will assume below that the
low-dimensional spin-glass quantum critical point does obey the
hyperscaling property (\ref{qab3}), even if the mean-field theory
for its particular situation does not.

We now consider the computations of the physical quantities
discussed in Section~\ref{sec:phys}. For the dynamic spin
susceptibility near the wavevector ${\bf K}$, note that symmetry
allows a coupling like
\begin{equation}
\zeta_4 \int d^2 r d \tau_1 d \tau_2 Q_{\alpha\beta}^{ab} ({\bf
r}, \tau_1, \tau_2) \mbox{Re}\left[\Phi_{x \alpha}^{a \ast} ({\bf
r}, \tau_1) \Phi_{x \beta}^b ({\bf r}, \tau_2)
\right],\label{qab5}
\end{equation}
along with the corresponding $(x \rightarrow y)$. Using the
expectation value (\ref{qab2}) in (\ref{qab5}) , we immediately
obtain a dynamic spin susceptibility of the form (\ref{D1}) with
\begin{equation}
M({\bf q}, \omega) \sim D(\omega) \label{qab6}.
\end{equation}
The singular part of the inverse susceptibility is ${\bf q}$
independent and a function of $\omega/T$, as in the local quantum
criticality scenario. Again, however, the scaling function in
(\ref{qab3}) is determined by a bulk $d$-dimensional quantum field
theory.

Finally, we turn to the electron self energy, and the associated
single particle and transport lifetimes. One approach, which
follows Ref.~\onlinecite{sg}, is to compute this perturbatively
from the coupling which follows directly from the identification
(\ref{qab})
\begin{eqnarray}
&& \zeta_5 \int d^2 r d\tau_1 d\tau_2 Q_{\alpha\beta}^{ab} ({\bf
r}, \tau_1, \tau_2) c_{\varsigma}^{a \dagger} ({\bf r}, \tau_1)
\tau^{\alpha}_{\varsigma \varsigma'} c_{\varsigma '}^a ({\bf r},
\tau_1) \nonumber \\
&&~~~~~~~~~~~~~~~\times c_{\upsilon}^{b \dagger} ({\bf r}, \tau_2)
\tau^{\alpha}_{\upsilon \upsilon'} c_{\upsilon '}^b ({\bf r},
\tau_2). \label{qab7}
\end{eqnarray}
Using (\ref{qab2}), we can easily compute the electron self
energy, $\Sigma$ perturbatively in $\zeta_5$, and at order
$\zeta_5^2$ we find the contribution
\begin{equation}
\Sigma ({\bf q}, \omega_n) \sim \zeta_5^2 T \sum_{\epsilon_n}
\mbox{sgn}(\epsilon_n) D (\omega_n + \epsilon_n) \label{qab8}
\end{equation}
which is momentum independent, but a singular function of
frequency and temperature. Using (\ref{qab3}), we deduce that
$\Sigma$ also obeys the scaling form
\begin{equation}
\Sigma ({\bf q}, \omega_n) \sim T^{\mu + 1} F_6 (\omega/T)
\label{qab9}
\end{equation}
with
\begin{eqnarray}
\mbox{Im} \left[F_6 (y)\right] &=& \int_0^{\infty} \frac{d
\Omega}{\pi} \mbox{Im} \left[ F_5 (\Omega) \right] \nonumber
\\&\times& \left( 2 n(\Omega) + f(\Omega-y) + f(\Omega+y)
\right),~~~~
\end{eqnarray}
a relationship similar to (\ref{fse5}). Note again the local
nature of the self energy. As the expression (\ref{qab8}) scatters
the electron by large momenta around the Fermi surface, there is
little difference between single-particle and transport scattering
rates, and the latter can be deduced directly from (\ref{qab9}).
It should also be noted here that the present perturbative
computation for $\Sigma$ is known to break down in certain cases:
this was pointed out by Parcollet and Georges \cite{gp} who showed
that the electron self energy was far more singular than
(\ref{qab9}) for the spin glass models of
Ref.~\onlinecite{sy2,gp,gps} which had $\mu=0$. Nevertheless, the
local nature of the self energy, and $\omega/T$ scaling was
preserved.

\section{Computations for RG equations}
\label{oneloop}

This appendix will present some further details of the derivation
of the renormalization group equations for Case B, described by
the action $\mathcal{S}_{xy}+ \mathcal{S}_I$. We noted in
Section~\ref{rgb} that the topological structure of the diagrams
is identical to those considered in Ref~\onlinecite{balents}, and
so we will not reproduce the diagrams here.

The exponents $\eta$ and $z$ arise from the momentum and frequency
dependence of the self energy $\Sigma_\phi$ of the $\phi_{1.2}$
field. To leading order, the only such contribution to $\Sigma$ is
\begin{eqnarray}
\Sigma_\phi (q, \omega ) &=& g^2 \int \frac{d^d p}{(2 \pi)^d}
\frac{d \Omega}{2 \pi}
 \frac{|{\bf p}|}{(|{\bf p}|+a_I |\Omega|)} \nonumber \\
 &~&~~~~~~\times \frac{1}{(({\bf p}+{\bf q})^2 + (\omega+\Omega)^2 +
 s)} \label{sen}
\end{eqnarray}
where $q = |{\bf q}|$. By an expansion of $\Sigma_\phi$ for small
$q$ and $\omega$, we find
\begin{eqnarray}
\eta &=&  \left. \left( - \frac{d}{dq^2} \right)
\Sigma_{\phi\Lambda} (q, 0)
\right|_{q=0} \nonumber \\
\eta + 2(z-1) &=&  \left. \left( - \frac{d}{d\omega^2} \right)
\Sigma_{\phi\Lambda} (0, \omega) \right|_{\omega=0}. \label{etaz}
\end{eqnarray}
Here the subscript $\Lambda$ is meant to indicate that the
integrals in (\ref{sen}) are to be carried out as part of a
renormalization group procedure. The precise nature of the
integral then depends upon the scheme being used:
\newline
({\em i}) in the dimensional regularization/minimal subtraction
scheme, we perform the integrals in arbitrary dimension $d$, and
pick out online the poles in $\epsilon=3-d$ in (\ref{etaz});
\newline
({\em ii}) in the fixed dimension approach \cite{parisi}, we
perform the integral in $d=2$ for a finite $s$, take the
derivative of the result with respect to $\sqrt{s}$, and set
$s=1$.
\newline
({\em iii}) in the soft-cutoff approach \cite{brezin,rome2}, we
perform the integrals with a smooth cutoff on the scale $\Lambda$,
take the $\Lambda$ derivative, and then set $\Lambda = 1$.
\newline
Each of these approaches yields closely related expressions for
$\mathcal{J}(a_I)$ and $\mathcal{K}(a_I)$.

The remaining terms in (\ref{rg1}) arise from simple one-loop
integrals which can be evaluated at zero external momenta and
frequency; these yield the expression
\begin{equation}
\mathcal{I}(m,a_I) =  \int_{\Lambda} \frac{d^d p}{(2 \pi)^d}
\frac{d \Omega}{2 \pi} \frac{|{\bf p}|^m}{(|{\bf
p}|+a_I|\Omega|)^m (p^2 + \Omega^2 + s)^2}.
\end{equation}
Again the $\Lambda$ subscript indicates that the integrals are to
be evaluated in one of the renormalization group schemes discussed
below (\ref{etaz}).

We can now present our explicit results for $\mathcal{I}(n,a_I)$,
$\mathcal{J} (a_I)$, and $\mathcal{K} (a_I)$, for the respective
schemes.

In the dimensional regularization approach, picking poles in
$\epsilon=3-d$, we obtained
\begin{eqnarray}
\mathcal{I} (0,a) & = & \frac{1}{8 \pi^2}\nonumber \\
\mathcal{I} (1,a) & = & \frac{4 a^3 \ln a -2 a^3 + 3 \pi a^2
- 2a + \pi}{8 \pi^3 (1+a^2)^2} \nonumber \\
\mathcal{I} (2,a) & = & \frac{4 a^5 - 3 \pi a^4 + 16 a^3 \ln a
+ 6 \pi a^2 - 4a + \pi}{8 \pi^3 (1+a^2)^3}\nonumber \\
\mathcal{J} (a) & = & \frac{(3-2\ln a)a^5 - 3 \pi a^4 + (2 + 6 \ln a)a^3
+ \pi a^2 -a}{12 \pi^3 (1 + a^2)^3}\nonumber \\
\mathcal{K} (a) & = & -3 \mathcal{J} (a)
\end{eqnarray}
where we have dropped the subscript $I$ on $a$ for brevity.

Similar expressions were obtained in the fixed-dimension approach
in $d=2$:
\begin{eqnarray}
\mathcal{I} (0,a) & = & \frac{1}{8 \pi}\nonumber \\
\mathcal{I} (1,a) & = & \frac{(1-a) \sqrt{1+a^2} + a^2
\mathcal{L} (a)}{8 \pi (1+a^2)^{3/2}} \nonumber \\
\mathcal{I} (2,a) & = & \frac{(a^3 - 2 a^2 - 2 a + 1)
\sqrt{1+a^2} + 3a^2 \mathcal{L} (a)}{8 \pi (1+a^2)^{5/2}}\nonumber \\
\mathcal{J} (a) & = & \frac{a(2a^2 - 3 a -  1)
\sqrt{1+a^2} - a^2 (a^2 -2) \mathcal{L} (a)}{16 \pi (1+a^2)^{5/2}}\nonumber \\
\mathcal{K} (a) & = & -2 \mathcal{J} (a)
\end{eqnarray}
where
\begin{equation}
\mathcal{L}(a) \equiv \ln \left( \frac{1 + a + \sqrt{1+a^2}}{1 + a
- \sqrt{1 + a^2}} \right)
\end{equation}

We also obtained results in scheme ({\em iii}) above, with both
soft and hard cutoffs; however we refrain from giving complete
expressions as the results are similar to those above.



\begin{thebibliography}{}
\bibitem{foot} We should note here a possible exception to this canonical weak
coupling scenario: Chubukov and co-workers \protect\cite{andrey}
have studied spin and charge density wave (SDW and CDW)
transitions in two dimensions, and have pointed out strong
coupling effects in the electron self energy and possibly other
observables.
\bibitem{commensurate} For commensurate wavevectors, the $S_1$
factor is reduced to the discrete space $Z_p$, but all of our
results continue to apply as long as $p>2$.
\end{thebibliography}
\end{document}